# Morphological Variation in a Toroid Generated from a Single Polymer Chain

**Y. Takenaka and K. Yoshikawa**

*Department of Physics, Graduate School of Science, Kyoto University, Kyoto 606-8502, Japan*

**Y. Yoshikawa**

*Department of Food and Nutrition, Nagoya Bunri College, Nagoya 451-0077, Japan*

**Y. Koyama**

*Department of Textile Science, Otsuma Women's University, Tokyo 102-8357, Japan*

**T. Kanbe**

*Division of Molecular Mycology and Medicine, Department of Advanced Medical Science, Center for Neural Disease and Cancer, Nagoya Graduate School of Medicine, 466-8550, Japan*

**Abstract**

A single semiflexible polymer chain folds into a toroidal object under poor solvent conditions. In this study, we examined the morphological change in such a toroidal state as a function of the width and stiffness of the chain together with the surface energy, which characterizes the segmental interaction parameter. Changes in the thickness and outer/inner radius are interpreted in terms of these parameters. Our theoretical expectation corresponds to the actual morphological changes in a single giant DNA molecule as observed by electron microscopy.

**INTRODUCTION**

It is well known that a toroid structure is generated from a giant DNA molecule upon the addition of condensing agents such as multivalent cations, polyamines and neutral flexible polymers[1-8]. After a full survey of previous experimental studies on a DNA toroid, Bloomfield stated that "--- the major diameter of the condensed particle remains almost constant on the order of 60-80 nm and does not seem to depend upon the length of an individual DNA molecule, so long as the DNA length is from 400 to over 40,000 base pairs (bp)."[9] In contrast, it has been reported that a giant toroid larger than 150 nm in diameter is generated under conditions in which the attractive interaction between segments is weak[3]. Several theoretical and experimental reports on the stability of a toroid have been published[10-19]. For example, Ubbink and Odijk[11] calculated the energy of a toroid by changing the surface tension, or the attractive potential between segments, under the framework of a constant toroidal volume. They showed that the major diameter increased and the thickness, or minor diameter, decreased accompanied by a decrease in the surface tension. Park et al.[12] theoretically examined the effect of disordered packing on a toroid by analyzing the defect energy. Vasilevskaya et al.[13] reported a mean-field approach on the stability of a toroid, with emphasis on the transition between an elongated coil and a compact toroidal state. Miller et al.[14] compared the free energy of a toroid with a hollow sphere, and found that the free energy of a hollow sphere becomes lower than that of a toroid accompanied by the increase in the characteristic parameter $\alpha$, which was first introduced by Ubbink and Odijk[11].

It has become clear that the conformational transition of a single semiflexible polymer from a coil into a toroidal structure can be interpreted in terms of the first-order phase transition between disordered and ordered states[20]. This phase transition exhibits unique characteristics as a finite system, and shows that a toroidal structure may be destabilized when the length of the polymer chain becomes infinitely long. Thus, the surface energy plays a significant role in determining the stable structure of a folded compact state in a single semiflexible chain.

In the present study, we examined the morphological change in a toroidal structure. We studied the effects of physico-chemical properties such as the stiffness and thickness of a chain with an approximate analytical treatment. Semi-quantitative agreement with actual experimental results with a single giant DNA molecule is demonstrated.

**THEORETICAL ANALYSIS**

**(i) Method**

Figure 1 shows a schematic drawing of the setting for this analysis. We consider the upper half of a cross-section of a toroid under the assumption of axisymmetry. The shape of the cross-section of a toroid (x, y) can be approximated by the product of a line and a semicircle as follows;

$$y = (cx + d)\sqrt{r^2 - (x - R)^2} \quad (1)$$

where c, d, r and R are the toroidal parameters that characterize the toroidal morphology, as shown in Fig. 1. We adopt two effective property parameters for a polymer, physico-chemical parameters, normalized by the toroidal surface tension σ: the thickness D and the persistence length P of a polymer. Throughout this paper, we set σ=0.75. Based on this condition, the following equations are given for the bending energy of a polymer $F_b$, the surface energy of a toroid $F_s$, and the contour length of a polymer L:

$$F_b = \frac{4\pi P}{D} \int_{R-r}^{R+r} \frac{y}{x} dx \quad (2)$$

$$F_s = \sigma \cdot 2\pi X_0 \cdot l \quad (3)$$

$$L = 4\pi \int_{R-r}^{R+r} xy\, dx \quad (4)$$

where $X_0$ is the x-coordinate of the center of mass of the cross-section and *l* is the length around the cross-section. We consider a toroid formed from a single polymer chain with a constant contour length L. With this constraint, we can decrease the number of variables to three. Thus, we consider the effect of the physico-chemical parameters on the toroidal parameters, c, r and R, in the present study. The total energy F is given as

$$F = F_b + F_s \quad (5).$$

By minimizing equation (5) with respect to these three parameters, we can obtain the optimum dimension of a toroid.

**(ii) Result**

Figure 2 shows the relationship between each physico-chemical parameter, D and P, and the toroidal parameters, r, R and c. A compact toroid (c, r and R nearly equal to 0, 10 and 30 respectively) appears when D=1.58 and P=50, and this shows a typical toroidal shape that has been frequently reported in past studies[9]. Figure 2a shows the effect of D on r, R and c under P=50. With an increase in D, r and R increase, whereas c decreases. As depicted in the figure, with an increase in thickness D, the compact toroid becomes swollen and globular where the inner diameter becomes much smaller. In addition, the inner hole of the toroid disappears for sufficiently large D, where r equals R.

Figure 2b shows the effect of P on r, R and c when D=1.58.   With an increase in P, R and r increase.   This means that the inner diameter of a toroid increases with an increase in P.   For sufficiently small P, the inner hole of a toroid disappears, indicating that a toroid can not be formed.

Based on these results regarding the effects of D and P, it is possible to predict the formation of a giant toroid by adopting proper values of P and D, as shown in Fig. 2c.

**EXPERIMENTAL**

In this section, some of the above theoretical considerations are discussed in connection with the experimental results with toroids formed from the same giant DNA molecules.

**(i) Method**

T4 phage DNA, 166 kbp with a contour length of 57 μm, was purchased from Nippon Gene (Toyama, Japan).   Maltosylated amino-pendant polyethylene glycol (Mal-PEG-A) was also used.   Samples A were prepared by adding 0.6 mM spermidine to 0.3 μM DNA solution.   Samples B were prepared by adding Mal-PEG-A to 0.6 μM

DNA solutions. They were mounted on carbon-coated copper grids (200 mesh), negative-stained with 1 % uranyl acetate, and observed with a JEOL 1200EX transmission electron microscope (TEM; Tokyo, Japan) at 100kV.

**(ii) Result**

Figure 3 shows the transmission electron microscopic (TEM) images of T4 DNA. Figure 3a shows the shape of a compact toroid obtained with sample A. Figure 3b shows a globular toroid with sample B, where DNA is folded with a relatively large excluded volume using a condensing agent. In other words, the effective thickness of a DNA chain is expected to be markedly larger than that of usual double-stranded DNA. Figure 3c shows a giant toroid formed at a high spermidine concentration (6.0 mM). Past studies on DNA compaction with polycations suggest that the attractive interaction between DNA segments becomes weaker at such a high concentration of spermidine[3], which corresponds to a decrease in surface tension $\sigma$.

**DISCUSSION**

We have advocated the minimal model to clarify the relationship between polymer properties and toroidal morphology. This model offers two merits. First, we can determine the shape of the cross-section of a toroid. Next, from an experimental point of view, we can determine the effect on DNA of a chemical added to the DNA solution by considering the image of a toroid; by determining whether the chemical effectively changes the thickness or stiffness of a DNA chain.

In this paper, we did not account for electronic interaction, including the effect of counter ions, in an explicit manner in the analysis of the stability of a toroidal structure. It may be important to note that the large morphological change in a toroid, as in Fig. 3, can be reproduced even without considering such an electronic effect. However, a unique structure such as interchain segregation has been found to be generated through the effect of a surviving negative charge[21, 22]. In the present paper, we studied the morphological variation of a toroid assuming a stable form as a toroid. When we consider the kinetics of the folding process of a polymer, other structures such as a rod will appear with the same property parameters[23, 24].

**ACKNOWLEDGEMENT**

This work was partly supported by the Grants-in-Aid for the 21st Century COE "Center for Diversity and Universality in Physics" and for Scientific Research in Priority Areas "Chemistry of Biological Processes Created by Water and Biomolecules" from the Ministry of Education, Culture, Sports, Science and Technology (MEXT) of Japan.

Figure 1. Schematic illustration of a toroid. The center of the toroid is set as the origin. The distances along the x-axis and the radius are r and R, respectively.

Figure 2. Result of the theoretical analysis. (a, b) Relation between the toroidal parameters and the effective property constants. The left vertical axes represent the thickness (r) and average radius (R) of a toroid and those on the right show a deformation parameter c. In each graph, there are two characteristic toroids; (a) Compact toroid (D=1.58 and P=50) and Globular toroid (D=13.7 and P=50) and (b) Compact toroid the same as Fig. 2a and a Fat toroid (D=1.58 and P=15). (c) Giant toroid (D=43 and P=2500).

Figure 3. TEM images of toroidal structures formed from single T4 DNA molecules. (a) Compact toroid in 0.6mM spermidine. (b) Globular toroid induced by Mal-PEG-A. (c) Giant toroid in 6mM spermidine (Reprinted in part with permission from reference 3. Copyright 1999 American Chemical Society). The scale bar is 100nm.

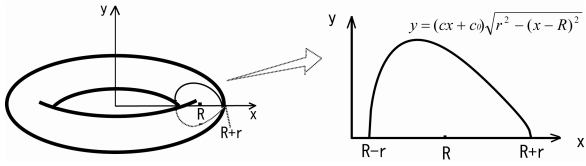

Fig. 1

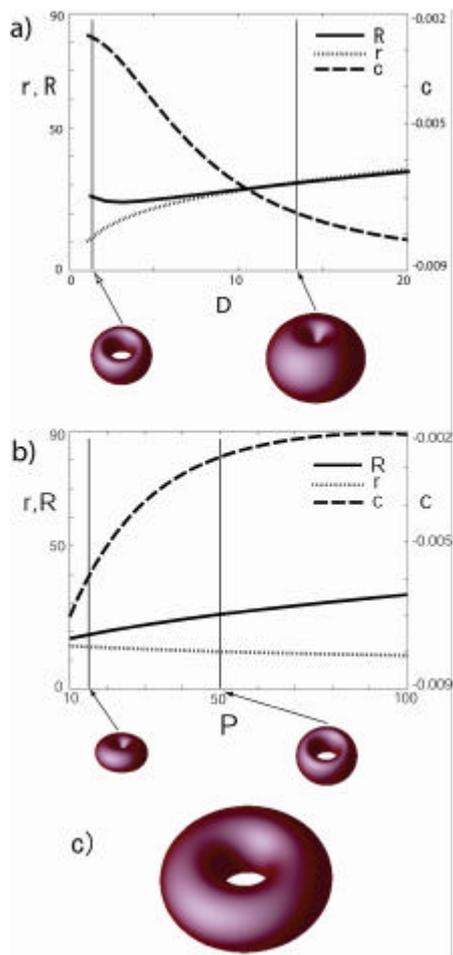

Fig. 2

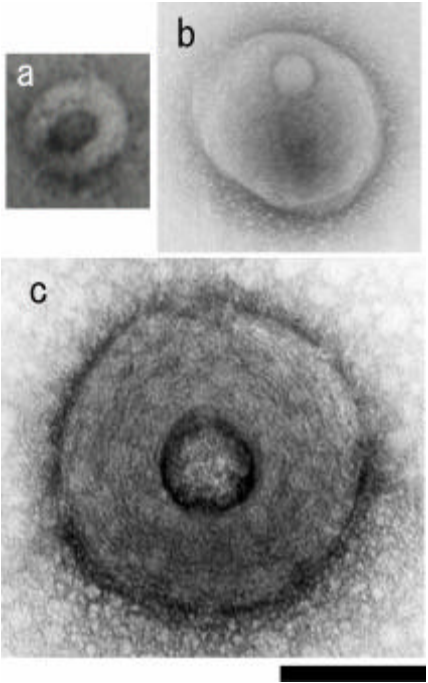

Fig. 3